# On-Demand Activation of Photochromic Nanoheaters for High Color Purity 3D Printing

**Authors:**   Alexander W. Powell[1], Alexandros Stavrinadis[1], Sotirios Christodoulou[1], Romain Quidant[1,2] *, Gerasimos Konstantatos[1,2] *

[1]ICFO-Institut de Ciencies Fotoniques, The Barcelona Institute of Science and Technology, 08860 Castelldefels (Barcelona), Spain

[2]ICREA- Institució Catalana de Recerca i Estudis Avançats, 08810 Barcelona, Spain.

* Corresponding authors: gerasimos.konstantatos@icfo.eu, romain.quidant@icfo.eu

**Abstract**

The creation of white and multicoloured 3D-printed objects with high colour fidelity via powder sintering processes is currently limited by discolouration from thermal sensitizers used in the printing process. Here we circumvent this problem by using switchable, photochromic tungsten oxide nanoparticles, which are colourless even at high concentrations. Upon ultraviolet illumination, the tungsten oxide nanoparticles can be reversibly activated making them highly absorbing in the infrared. Their strong infrared absorption upon activation renders them efficient photothermal sensitizers that can act as fusing agents for polymer powders in sintering-based 3D printing. The $WO_3$ nanoparticles show fast activation times, and when mixed with polyamide powders they exhibit a heating-to-colour-change ratio greatly exceeding other sensitizers in the literature. Upon mixing with coloured inks, powders containing $WO_3$ display identical colouration to a pristine powder. This demonstrates the potential of $WO_3$, and photochromic nanoparticles in general as a new class of material for advanced manufacturing.

**Keywords :** 3D-Printing, photochromism, photothermal, tungsten oxide, laser sintering, plasmonic

In recent years a new generation of composite materials with advanced functionalities has arisen, opening up dynamic new frontiers in both conventional and advanced manufacturing by exploiting the unique properties of nanomaterials. The earliest example of this for 3D printed materials was the creation of nanocomposites through blending polymers with ceramic nanoparticles (NPs) such as alumina[1] and nanoclays[2] to improve the mechanical properties of the printed objects. More recently, carbon nanomaterials such as carbon nanorods[3], graphene[4], and core-shell structures[5] have been used both to improve mechanical performance of 3D prints via laser sintering and to produce electrically conductive objects[6–8,10] via a variety of methods Silver nanoparticles have also been used to make antibacterial 3D prints[9], gold nanoinks have been used to make 3D printed electrochemical reactors for catalysis[10], and piezoelectric actuators and sensors have been printed with $BaTiO_3$ NP-polymer nanocomposite[11,12].

Another area in which nanomaterials have shown potential to disrupt current additive manufacturing (AM) technologies is in control over the aesthetic properties of the objects. Traditional 3D prints are printed in a monochrome and must be painted or dyed afterwards. There has been some experimentation with grading powders of different colour to produce a 'two tone' effect[13] but this is a cumbersome process and more recent work has focused on inkjetting coloured inks into the powder during the printing process along with a thermal sensitizer using the 'High Speed Sintering' approach developed by Hopkinson et al[14]. However the colour fidelity of these prints is strongly dependent on the properties of the sensitizers used. Gold nanorods (GNRs) were shown to operate as a photothermal sensitizer to allow the rapid sintering of polymer powders into 3D objects using low-power light sources, and without significantly altering the aesthetic properties[15]. This was achieved by tuning the resonant absorption of the GNRs into the near infrared (NIR), leaving only weak absorption in the visible. Whilst this work showed excellent results, there are some drawbacks to using gold nanorods as a photothermal sensitizer : GNRs are not stable at the very high temperatures required to sinter new high-performance polymers such as PEEK[16] (melting temperature of 343 C[17]), but crucially GNRs intrinsically have weak absorption within the visible spectrum, which impacts on the colouration of prints when used in high concentrations, imposing intrinsic limits to the amount of GNRs one can use without affecting the color purity of the printed objects. In fact, this can be a problem for almost all plasmonic absorbers used for photothermal studies, in view of their absorption tail in the visible that becomes significant when the NPs are present in large quantities[18–24].

In order to reach high color purity in 3D prints, we posited that instead of using permanent absorbers to generate heat, the use of switchable absorbers would allow us to print objects with very high color purity and whiteness. This is important as a strongly coloured powder will limit the colour gamut available for a print, which can impact negatively on both the aesthetics and applications such as barcoding of prints[25]. These would only possess strong optical absorption properties when they are needed to facilitate photothermal sintering, and would otherwise be transparent.  Essentially, these switchable absorbers would act as programmable elements, which during the 3D printing process would be programmed, preferably optically, to switch ON and function as photothermal sensitizers, while later on they would switch OFF. To this end, we sought the use of photochromic materials, whose absorption can be altered via an external optical stimulus. In particular, we chose inorganic oxide semiconducting NPs due to their inherent stability at the elevated temperatures required in 3D printing[26].

One of the most promising of these materials is tungsten (VI) oxide, $WO_3$, which has a strong absorption in the NIR that can be efficiently switched on and off using electricity (electrochromic)[21] or ultraviolet radiation (photochromic)[27]. Moreover, $WO_3$ is stable at high temperatures well above 400 C[26,28], and can be readily synthesized as nanoparticles using cheap reactants[29,30]. In this work, we utilize for the first time the photochromic properties of $WO_3$ NPs to develop a 3D printing method that offers very high color purity and achieves white 3D printed objects via photothermal sintering with low cost NIR light sources.

To achieve this we synthesized $WO_3$ uncoated nanoparticles and mixed them with commercial AM powders to create a photochromic nanocomposite powder. We characterized the photochromic properties of these particles in ways not before applied to $WO_3$ NPs, and their behavior is found to agree with the general model of excitation seen in previous studies. We demonstrate the strong, rapid photochromic response of the $WO_3$ particles, and compare the photothermal heating and colouration of the powders to equivalent composite powders made with the industry standard carbon black (CB) as well as recently reported plasmonic nano-heaters based on GNRs. We demonstrate the sintering of these powders via a low powder laser (a 1 W diode laser, compared to the ~40 W $CO_2$ lasers in traditional SLS machines[31]) to make 3D printed objects in order to demonstrate the colour contrast between the three sensitizers. Finally, we compare effect of the NP heat-sources on the hues of powders when mixed with coloured inks, highlighting the excellent utility of $WO_3$ for 3D printing white and finely coloured objects.

The photochromic effect of Tungsten oxide is due to the formation and decay of a hydrogen tungsten bronze ($HWO_3$) created by photo-induced interactions with the surrounding environment[32]. Briefly, the photo-generated electrons induced by UV-illumination decompose the water molecules on the surface of the NCs while reducing tungsten to $W^{+5}$ oxidation state forming

HWO$_3$ near the surface, creating a doped hydrogen bronze[27,32,33]. Upon illumination, these extra electrons in a WO$_3$ nanoparticle can interact with light in one of two ways. (i) Charge hopping between lattice sites of different W oxidation levels (the small polaron model)[32,34,35], (ii) absorption by free electrons and the excitation of surface plasmons[34–36]. The predominance of each mechanism is dependent on the substoichiometry of the particle and the level of hydrogen doping at any given time[34,37]. This process has been shown to be highly reversible and stable over thousands of cycles[38].

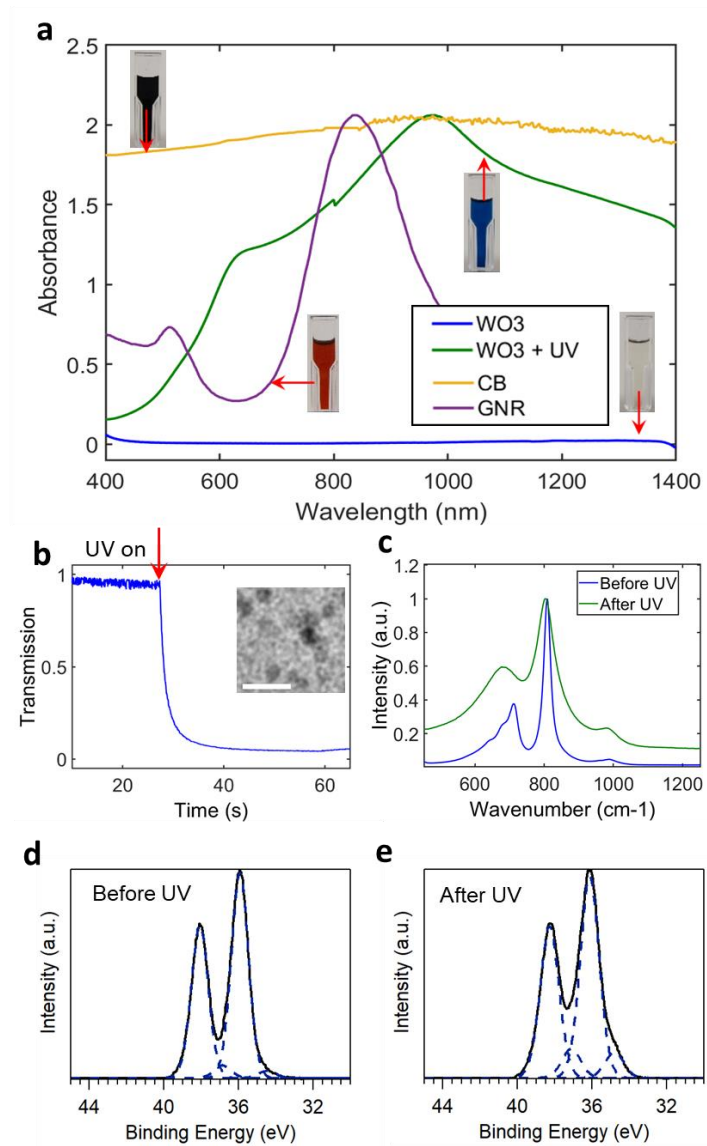

**Figure 1:** a) Absorption plots of GNRs, Carbon black, and $WO_3$ nanocrystals before and after UV illumination for 20 s. Photos of vials containing concentrated solutions of each material are also shown for clarity. b) The transmission of an 808 nm laser through a $WO_3$ solution while the samples is excited with UV light. The Inset shows a TEM image of the $WO_3$ NPs, the scale bar shows 20 nm. c) Raman spectra of the particles before and after illumination. XPS spectra from W 4f core level of $WO_3$ NCs before (d) and after illumination (e).

$WO_3$ NPs are synthesized following the procedure outlined by Brutsch et al[29] and were dispersed in water at a concentration of 3 g/L and their spectra are shown in Fig. 1a along with solutions of CB and GNRs, next to photos of the solutions. For WO$_3$ the spectra is shown in the relaxed state, and the photoexcited state after illumination with ultraviolet (UV) light for 20 seconds. Whilst the carbon solution is completely black, the GNR and excited WO$_3$ spectra show a strong absorption band focused on a resonant value in the NIR. Dilute GNR solutions appear mostly transparent, but at high concentrations they have a distinct blood-red colour. The relaxed state of the WO$_3$ shows negligible absorption until the bandgap at wavelengths shorter than 400 nm, and the solution appears transparent in the visible, even at high concentrations. Upon UV photoexcitation the WO$_3$ shows strong absorption in the NIR with a tail in the visible strong enough to turn it a deep blue. This is a fully reversible and repeatable process, and after each excitation the sample reverts to its original colourless state. This effect occurs in dry samples as well as solutions (Figs S3, S4 & S6). In all cases environmental factors play an important role in determining the strength and speed of the photochromic response.

There are two distinct peaks present at 640 nm and 974 nm. Several previous studies have observed this double peak feature and have attributed the long wavelength peak to a plasmonic resonance in the nanoparticle, and the shorter wavelength peak to polaron hopping, although the exact position of peaks has been found to be highly dependent on the size and composition of the nanoparticles[34,35]. Thus the potential of this material is already clear, if the strong absorption peak shown in Fig 1a can be switched on and off, then it is possible to achieve strong heating and sintering of powders via IR absorption, and later the absorption will switch off to leave the printed object uncoloured.

Fig. 1b shows the transmission dynamics of an 808 nm laser through a vial of 1 g/L WO$_3$ in H$_2$O, before and after excitation via a UV light, demonstrating the speed of this reaction. The transmission is reduced to 25% of its start point in, < 2 seconds and saturates at < 5% after 30s.

This demonstrates that the photochromic effect is excited very rapidly in the particles, and is a process compatible with 3D printing manufacturing. The inset of Fig 1b shows a TEM image of the $WO_3$ NPs. Several images were taken and a survey of NPs found their average size to be $6 \pm 2$ nm (See Fig. S2).

The origin of photochromism in $WO_3$ as described earlier is associated to its crystal structure and chemical composition being altered upon UV illumination in the presence of a hydrogen donor. We therefore performed Raman and X-ray Photoelectron spectroscopy to verify the mechanism at play in our NPs. Raman analysis in Fig. 1c shows two sharp peaks around 712 cm$^{-1}$ and 809 cm$^{-1}$, both resulting from O–$W^{6+}$–O stretching modes[32]. Upon UV illumination the peaks become significantly broader, which points to a decrease in WO3 crystallinity, which is typical of the formation of the defect $W^{5+}$ states in $WO_3$ and has been observed previously for electrochromic[39] and plasma-based excitation[40] of $WO_3$. We further confirm the photo-induced reduction of the tungsten by probing the W 4f core levels of the $WO_3$ NCs with X-ray photoelectron spectroscopy. In Fig. 1 d & e we show the W 4f core level spectra before and after UV illumination. The XPS spectra of the pristine $WO_3$ consisting of doublet peak due to spin-orbit coupling assigned as W $4f_{7/2}$ and W $4f_{5/2}$ with binding energies of 35.9 eV and 38.1 eV respectively, which corresponds to $W^{+6}$. Furthermore, the XPS analysis of the W4f spectra after UV illumination suggest the reduction of the $W^{+6}$ to $W^{+5}$ which appears as an additional doublet peak shifted 1eV at lower energies in line with the previous reports studying plasma excitation.[29] We have not observed a complete reduction of the tungsten into $W^{+5}$ oxidation state after UV-treatment which confirms that the photo-reduction mechanism takes place near the surface of the NCs as previously reported and agrees with the model of the formation of tungsten bronzes being responsible for the photochromic effect.[30]

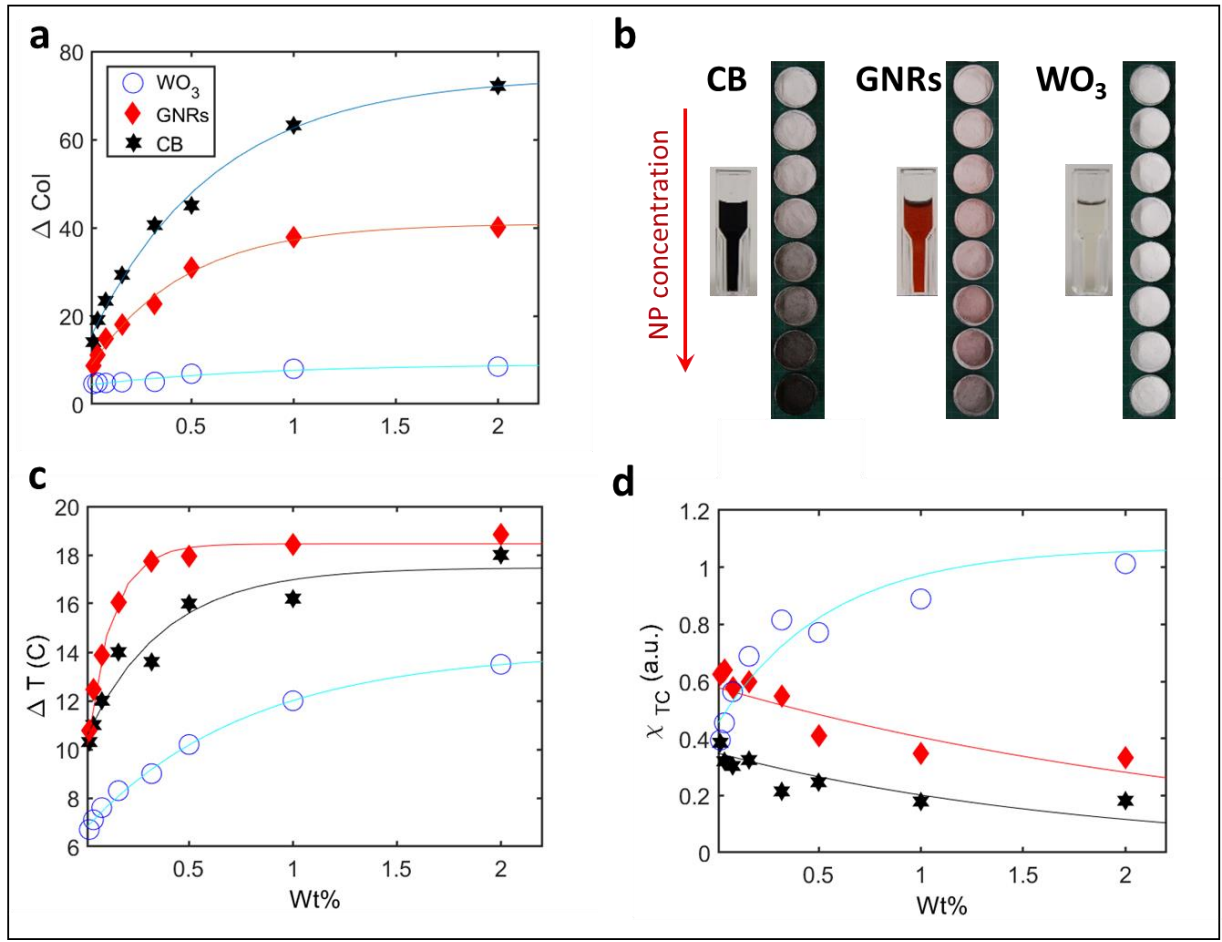

**Figure 2:** a) Colouration parameter *ΔCol* as a function of NP concentration for WO$_3$-PA12, CB-PA12 and GNR-PA12 powder mixtures. b) Photographs of the different concentration powders used highlighting the effect of each nanoparticle on the colour. c) The heating, ΔT (increase in temperature from ambient values) for each concentration of powder under illumination from a 940 nm 100W LED array for CB and WO$_3$ and an 810 nm 100W LED array for GNR samples. The heating figure for the WO$_3$-PA12 is given for the tungsten oxide in its active state after UV photoexcitation. d) The heating-to-colour-change ratio for each powder as a function of photothermal sensitizer concentration. Fit lines have been added as a visual aid.

The two key parameters for any photothermal sensitizer considered for 3-D printing are the degree to which they can enable the heating of a powder and the extent to which they change the colour of said powder and the final printed objects. In Figure 2a, the colour change from the pristine state, $\varDelta Col$, caused by the addition of $WO_3$, CB and GNR NPs to polyamide (PA12) powder with is shown as a function of NP concentration. Colouration is defined as $col = \sqrt{(a^*)^2+(b^*)^2 + (L^*)^2}$ from the CIEL*a*b* scale, (See Fig. S10) and so $\Delta Col = \sqrt{(\Delta a^*)^2+(\Delta b^*)^2 + (\Delta L^*)^2}$, with the base values of a*, b* and L* are taken from pristine PA12. High values of $\varDelta Col$ indicate a large change in the colour of the powder. It can be seen that even a tiny amount of CB has a marked impact on the powder colour - the addition of 0.02 % wt. of the powder leads to $\varDelta Col$ = 14, whereas the $WO_3$ has a negligible impact even at high concentrations, with a maximum value of $\varDelta Col$ = 8.5 for a 2 % wt. nanocomposite - 60 % the change in colour for 100x the sensitizer. This is illustrated visually in Fig. 2b, which shows that the $WO_3$-PA12 powder remains white in its relaxed state for all concentrations used. Fig. S3 shows that as with the solutions in Fig. 1a, the WO3-PA12 powders and solid melted samples also take on a bluish colour after excitation with UV light, which then decays back to white after relaxation. GNR's have a smaller impact on the colouration than CB, but significantly more than $WO_3$ – and the $\varDelta Col$ values lie almost at the midpoint of those of CB and $WO_3$ for all concentrations measured.

When illuminated at resonance with a 100 W LED array, the CB and GNRs heat more effectively than the $WO_3$ (in its active state after UV photoexcitation) at all concentrations, although the difference is less marked at higher concentrations, as shown in Fig. 2c. The GNRs have the strongest heating out of the three for all measured values. However the most important figure of merit is the heating that can be achieved for a change in colouration. The ideal photothermal sensitizer for colour 3DP will maximize heating and without altering the colouration. We therefore propose a new parameter, the heating-to-colour-change ratio, defined as : $\chi_{TC}=\varDelta T/\varDelta Col$ . Fig. 2d shows that for CB, with increasing concentration, the change in colouration is stronger than the increase in heating, whereas for $WO_3$, the change in colouration is negligible and so the increase in heating dominates. The benefit of the photochromicity of $WO_3$ is clear here, as in the excited state they show a significant photothermal heating, but once this state has decayed they are almost colourless, producing the high $\chi_{TC}$ oserved. The GNR's follow a similar trend to the CB, although with much higher $\chi_{TC}$ values. Very low concentrations of GNRs appear white and their very high heating allows for them to outperform low concentrations of $WO_3$, but as seen in Fig. 2c the powder quickly darkens and takes on a pinkish tinge with increasing concentrations, and its heating-to-colour-change ratio is quickly overtaken by WO3 for concentrations above 0.08 %wt. Powders with concentrations of GNRs greater than 0.02 %wt were found to have a pinkish tinge, especially notable when sintered, as seen in Fig. 3c.

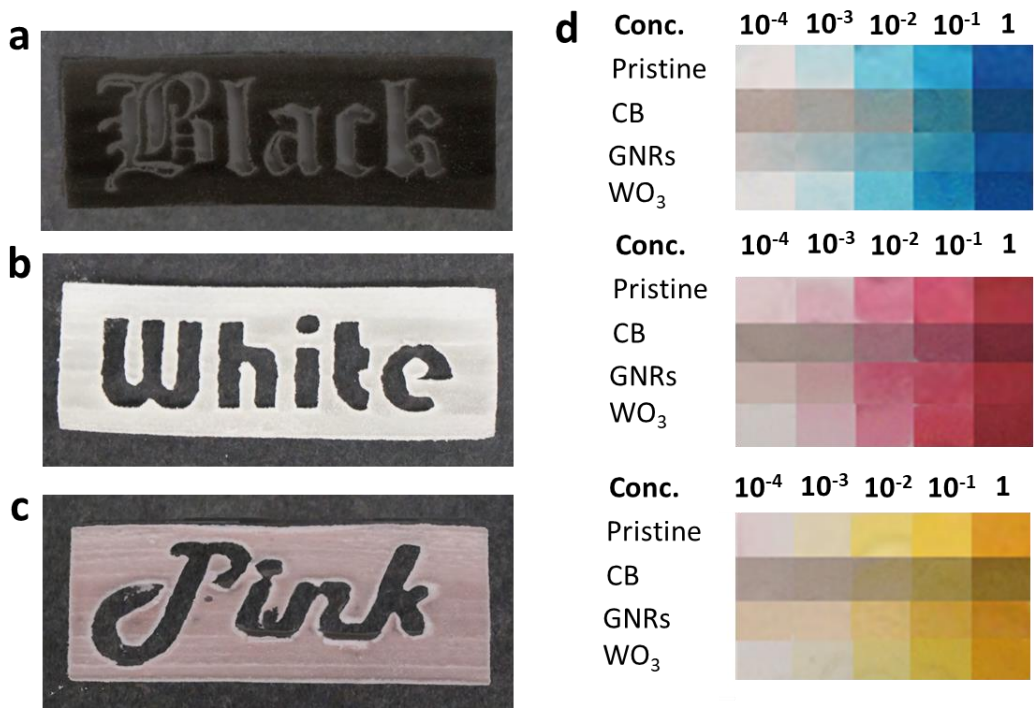

**Figure 3:** Laser sintered samples produced with 0.08 % wt. CB-PA12 powder (a), 1 % wt. WO3-PA12 powder (b), and 0.03 % wt. The samples are sintered rectangles with hollow letters. GNR-PA12 powder (concentrations chosen to give equivalent heating). (d) Photographs of different

concentrations of coloured ink dropped onto each of the three powders, highlighting the effect of the NP absorption on each primary colour.

We can therefore state that $WO_3$ is a more effective sensitizer than CB and GNRs in terms of the $\chi_{TC}$ parameter. Figure 3 illustrates the importance of this effect. A custom-made laser sintering bed, using a 1W 808 nm laser was used to sinter $WO_3$-PA12, CB-PA12 and GNR-PA12 nanocomposite powders in exactly the same method that would be used in an SLS 3D printer (Fig. S9). The $WO_3$-PA12 samples were illuminated via a UV lamp during sintering in order to maintain the highly absorbing excited state. The powders used to make these samples were mixed to have the same photothermal response at 808 nm, so in Fig. 3a the $WO_3$ has a concentration of 1 % wt, the CB has a concentration of 0.08% wt. and the GNRs have a concentration of 0.03 % wt in the powder. This shows the importance of considering both heating and colouration together. As the photothermal properties of the samples at the laser wavelength are identical, they will show the same $\Delta T$ for a given illumination, but the colour contrast between them is stark – the CB sample appears almost completely black and the GNR sample appears pink, whilst the $WO_3$ sample remains white.

The aesthetic impact of the photothermal sensitizers is also key for printing fully coloured objects, as the colour of the powder will obviously affect the colour of the finished prints. It is important to be able to print the full gamut of colours, as well as just in monochrome. This is investigated in Fig. 3d, which shows photos of PA12 powder in its pristine form and mixed in with $WO_3$, CB and GNRs at the same concentrations as previously with various concentrations of red, blue and yellow inks dropped onto the powder. The ink was used as received for the most concentrated example, and then diluted by a factor of 10 for each subsequent step, up until $10^{-4}$. Their coloration was then measured with a colorimeter and plotted on the CIEL*a*b* scale.

Evidently, there is very little difference between the colouration of the inks on $WO_3$ and pristine powders. This is reflected in the *ΔCol* vs. ink concentration plots (See Fig. S11), where for every ink the traces from the $WO_3$ and the pristine powder are basically indistinguishable. For the CB powder however, it is clear from the photos in Fig. 3d that the colours are all darker and not as well defined as with the $WO_3$. The *ΔCol* vs. concentration and L*a*b* plots (Fig. S11 & Fig. S12) show a significant difference between the colours on the CB powder compared to the others, and the lighter shades are simply not accessible due to the darkness of the powder. For the GNR's the stronger shades are almost indistinguishable from a pristine powder, but the lighter shades appear darker and redder, which would lead to a reduction of accuracy for lighter colours in printed objects. Thus the $WO_3$ powder shows a negligible change in hue compared to a pristine powder, even for very weak concentrations of inks, whereas the CB powder significantly restricts the available gamut to darker shades and the GNR powder reduces the accuracy of lighter shades.

In conclusion we have demonstrated that the use of photochromic tungsten oxide nanoparticles as a photothermal sensitizer has the potential to revolutionize high color quality 3D printing. $WO_3$ particles are easy and cheap to manufacture, show fast activation times, and demonstrate a heating-to-colour-change ratio far superior to other available sensitizers. When added to coloured inks, $WO_3$ reproduces even very pale shades perfectly in comparison to pristine powders. This work opens the door to a whole new class of nanomaterials for advanced manufacturing.

# METHODS

Gold nanorods were synthesized in-house via the seed-mediated method of Nikoobackht & El-Sayed[42], and coated in silica following Gorelikov and Matsuura[43]. Carbon black powder was purchased from PlasmaChem. $WO_3$ was synthesized following the procedure of Brutsch et al[29] and dispersed in a concentrated form in deionized water. This method was selected as it produces small, uncoated nanoparticles with a high surface-to-volume ratio, which has been shown to be advantageous for maximizing the photothermal heating efficiency of nanoparticles[44–46], and will also aid the rate of photochromic excitation, which largely occurs at the particle surface[32,34,37]. The small particle size also introduces quantum confinement effects which blueshift the absorption band edge out of the visible, leading to more transparent NPs (in their unexcited state). All of these effects will lead to a greater heating-to-colour-change ratio for the $WO_3$-PA12 composites, which is highly desirable

Absorption measurements were obtained from particles in solution using a Cary 5000 spectrometer. TEM measurements were carried out in a JEOL JEM-2100 LaB6 transmission electron microscope operating at 200kV. Raman spectra were collected using a Renishaw InVia MicroRaman spectrometer, exciting the samples with a 100 mW diode laser at $\lambda=532$ nm, using a $\times 50$ magnification microscope objective. XPS measurements were performed with a Phoibos 150 analyzer (SPECS GmbH) in ultra-high vacuum conditions using a monochromatic Kalpha x-ray source (1486.74eV). The photochromic excitation speed tests were carried out using a custom setup defined in Figure S7.

The NPs were diluted in ethanol then mixed with pristine PA12 powder from Advanc3Dmaterials at differing wt % concentrations then dried in an oven at 60 C overnight. After drying powders were sieved prior to printing. For measuring the effects of the NPs on the colour of powders, various concentrations of commercially available inks were drop-cast onto the powders and the colouration measured with a colorimeter.

To perform the comparative tests between the different sensitizers, the powder, small samples of powder in a shallow aluminium holder were illuminated by a 100W 810 nm or 100W 940 nm LED array from Shenzhen Hanhua Opto Co. The temperature of all samples was measured using a FLIR ax5 thermal camera, and the colour tests were performed using a PCE instruments PCE-CSM 4 colorimeter.

To sinter the printed shapes in Fig. 3, a custom made laser writing setup was used, as described previously[15]. Samples were preheated by a hotplate below and IR lamps from above in order to mimic the conditions in a SLS printer. The bulk of the powder was heated to 150C using the hotplate and then the top layer heated to 172.5 C using the lamps. A Dymax Bluewave 200 UV lightsource was used to illuminate the $WO_3$-PA12 sample during this process to maintain the nanoparticles in their photoexcited state. A 1W 808 nm diode laser from Thorlabs was used to sinter the powder and define the shape of the prints.

**Supporting Information**

Showing illustrations of the experiments used and additional results on the colouration of powders after the addition of the inks.

**Acknowledgements**

The authors acknowledge financial support from the European Community's Seventh Framework Program under grant QnanoMECA (64790), Fundació Privada Cellex, the program CERCA, the Spanish Ministry of Economy and Competitiveness, through the "Severo Ochoa" Programme for Centres of Excellence in R&D (SEV-2015-0522) and grant FIS2016-80293-R, the European Research Council (ERC) under the European Union's Horizon 2020 research and innovation programme (grant agreement No 725165), the Spanish Ministry of Economy and Competitiveness (MINECO) and the "Fondo Europeo de Desarrollo Regional" (FEDER) through grant TEC2017-88655-R.

**ToC**

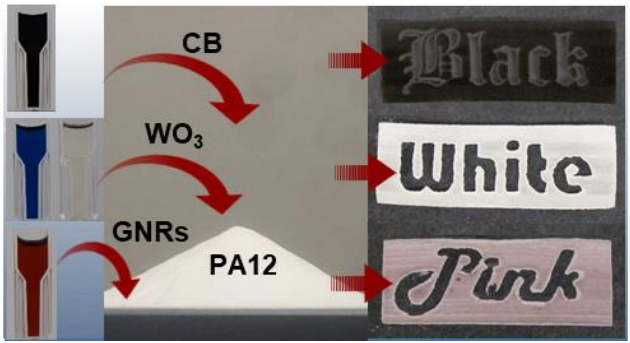